\documentclass[twocolumn]{aastex631}
\pdfoutput=1

\newcommand\kms{km s$^{-1}$}
\newcommand\masyr{mas yr$^{-1}$}

\usepackage{threeparttable,booktabs,soul}
\usepackage{longtable}

\begin{document}

\title{Kinematic Age of the $\beta$-Pictoris Moving Group}

\author[0000-0003-3651-2924]{Jinhee Lee}
\altaffiliation{KASI-Arizona Fellow}
\affiliation{Korea Astronomy and Space Science Institute (KASI), 776 Daedeok-daero, Yuseong-gu, Daejeon 34055, Republic of Korea}
\affiliation{Steward Observatory, The University of Arizona, 933 N. Cherry Ave, Tucson, AZ 85721, USA}
\email{jinhee@kasi.re.kr}

\author[0000-0002-5815-7372]{Inseok Song}
\affiliation{Department of Physics and Astronomy, The University of Georgia, Athens, GA 30602, USA}



\begin{abstract}
Accurate age estimation of nearby young moving groups (NYMGs) is important as they serve as crucial testbeds in various fields of astrophysics, including formation and evolution of stars, planets, as well as loose stellar associations. The $\beta$-Pictoris moving group (BPMG), being one of the closest and youngest  NYMGs, has been extensively investigated, and its estimated ages have a wide range from $\sim$10 to 25 Myr, depending on the age estimation methods and data used.
Unlike other age dating methods, kinematic traceback analysis offers a model-independent age assessment hence the merit in comparing many seemingly discordant age estimates. 
In this study, we determine the kinematic ages of the BPMG using three methods: probabilistic volume calculation, mean pairwise distance calculation, and covariance matrix analysis. 
These methods yield consistent results, with estimated ages in the range of 14 to 20 Myr. Implementing corrections to radial velocities due to gravitational redshift and convectional blueshift increases the ages by $\sim2-4$ Myr. Conversely, considering data uncertainties decreases the estimated ages by 1 to 2 Myr. Taken together, our analysis determined the kinematic age of BPMG to be 16.3$^{+3.4}_{-2.1}$ Myr.
This age is significantly younger than the commonly accepted age of the BPMG ($\sim$24 Myr) determined primarily from the lithium depletion boundary analysis. This younger kinematic age may point to the discrepancy between the luminosity evolution and lithium depletion models or the presence of unaccounted systematic error in the method. This result underscores the necessity for systematic reevaluations of age-dating methods for nearby, young moving groups.
\end{abstract}



\section{Introduction} \label{sec:intro}
The $\beta$-Pictoris moving group (BPMG) stands out as one of the most extensively studied stellar groups due to its proximity and youth, as evidenced by various works (e.g., \citealt{son03, alo15, ell16, shk17, cro23}). When the mass function of the currently known BPMG members was compared to those of other stellar groups (e.g., $\alpha$Per, Pleiades, and Praesepe in \citealt{lod19}), it appears that the membership survey of BPMG is mostly complete down to $\sim$0.035 solar mass ($\sim$late T type, e.g., \citealt{lee24}). Consequently, BPMG serves as an ideal benchmark for studying the evolution of stellar groups, as well as the lowest mass stars and brown dwarfs. In addition, similar to other nearby young moving groups (NYMGs), BPMG members are prime targets for exoplanet direct imaging; well-known exoplanets discovered by direct imaging ($\beta$-Pictoris b and 51 Eri b; \citealt{lag10, mac15}) are members of the BPMG and they have become testbeds for studying the early evolutionary stages of wide-orbit massive planets \citep{mar08, nie13, bow16, bar19}.

%% The text below was copied from the 2023 ADAP/NYSARC proposal. Update references.
Most stars are born in clusters, with over 90\% expected to dissipate into the field within 1 Gyr \citep{kru19, lam06}. Young stars formed in a diffuse environment, such as NYMGs, are gravitationally unbound and can be spread on the sky plane over $100^2$ degrees or more. This is similar to stars in the Taurus star-forming region, which likely represents an earlier stage of the NYMG kinematic evolution. The fraction of stars in gravitationally bound groups (open and globular clusters) in the Milky Way is small ($\sim10\%$; \citealt{kru19}), which implies that most stars are formed in unbound groups like NYMGs. Gaia data, combined with ground-based high-resolution spectroscopic surveys that can establish radial velocities (RVs) with a precision of $<1$ \kms, provide the means to obtain detailed kinematic evolutionary histories for NYMGs. These kinematic fingerprints can reveal the dynamical mechanisms through which, and timescales over which, members of NYMGs disperse into the young field star population. %(ADD REFERENCES). 
Numerous structures of stellar populations, known as tails or streams, in the Milky Way were discovered recently \citep{mei21, tar22, bha22} with extensions over hundreds of parsecs in most cases. Only young stellar groups are detectable in (or as) such structures because they are dissipated in $\sim100$ Myr \citep{kru19}. Given their age range ($10-100$ Myr), NYMGs offer unique opportunities to study various evolutionary stages of sparse stellar groups so as to address several fundamental questions. 

Studying the kinematic evolution of NYMGs provides another piece of crucial information, i.e., age. The age of an ensemble of young stars such as an NYMG, can be determined using the theoretical isochronal fitting, lithium depletion boundary, IR-excess, and stellar activity indicators such as Ca H \& K lines, X-ray luminosity, NUV magnitudes, and H-$\alpha$ lines \citep{sod10}. Based on these methods, the age of BPMG has been estimated to be $\sim$12 to 25 Myr (e.g., \citealt{zuc01, mam14, bin14}). A dispersing stellar group can provide a model-independent age by determining the time when its members occupied the smallest volume in the past. This age-dating method, known as kinematic expansion or traceback age-dating, has been successfully applied to some young NYMGs \citep{duc14, mir18, gal23}. Generally, age estimates from the kinematic expansion method for BPMG show systematically younger ages ($12-20$ Myr; \citealt{ort02,son03,cru19, mir20,cou23}) compared to the widely accepted lithium depletion boundary age of $21-26$ Myr \citep{bin14, mal14}, prompting more systematic studies on the ages of NYMGs.

The wide range of different kinematic ages obtained for BPMG illustrates the sensitive dependence on input data and methods in calculating kinematic ages. Despite the precision of astrometric data (R.A., Dec., parallax, and proper motion) from Gaia and sub-km/sec RV precision, the propagation of uncertainty in position plays a major factor. For typical NYMG members, their positions in 3D space (e.g. XYZ in Cartesian coordinate) can be precisely calculated thanks to the superb precision of Gaia data. Therefore, uncertainties of traceback positions are generally limited by the precision of the RV measurements. For example, with a velocity uncertainty as small as 1 \kms, it contributes to a 1 parsec over 1 Myr of traceback calculation. For the upper age bound of BPMG at $\sim25$ Myr, the velocity uncertainty of \kms\ will result in a 25 pc uncertainty in traced back positions. In addition to the effect of uncertainty propagation, kinematic age dating results are sensitive to the input list of members and employed methods.

In this study, we derive the kinematic age of BPMG using three different methods to investigate the dependency of the resultant age on the method choice. 
In Sections \ref{sec:data} and \ref{sec:methods}, we will explain data and methods, respectively. Then, we will present the results and discussion in Sections \ref{sec:results} and \ref{sec:discussion}, respectively. The conclusion is presented in Section \ref{sec:conclusion}.

\section{Membership List and Input Data}
\label{sec:data}

\subsection{Member List \label{sec:memblist}}

\begin{figure}[htb!]
\centering
    \plotone{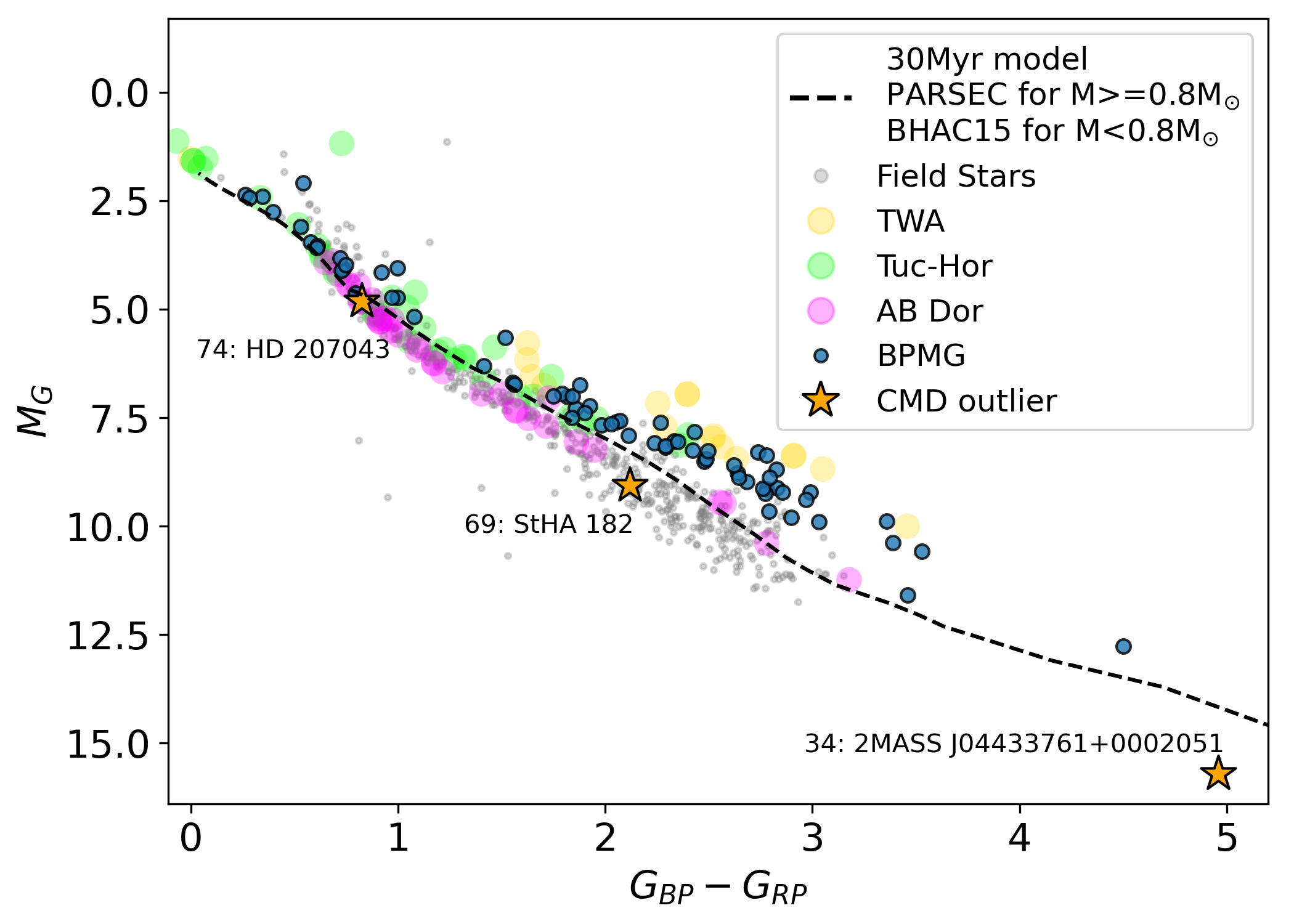}
    \caption{A color-magnitude diagram (CMD) of BPMG members is displayed, and input data are obtained from the Gaia DR3 catalog. Stars that meet the CMD criteria are indicated by blue circles, while those inconsistent with other BPMG members in the CMD are denoted as orange stars, with their corresponding star ID numbers and names referenced in Table~\ref{tab:members}. 
    Two theoretical isochrone models with an age of 30 Myr are displayed to show the youth of the members. The PARSEC model \citep{bre12} is displayed for the higher mass regime ($\geq0.8M_\odot$; AFG types), while the BHAC15 model \citep{bar15} is presented for the lower mass regime ($<0.8M_\odot$; KM types). Members of other NYMGs and field stars are presented for comparison (TWA: $\sim$8 Myr, Tuc-Hor: $\sim$30 Myr, and AB Dor: $\sim$100 Myr).}
    \label{fig:cmd_outliers}
\end{figure}

\begin{figure}[htb!]  
\centering
   \includegraphics[width=0.85\linewidth]{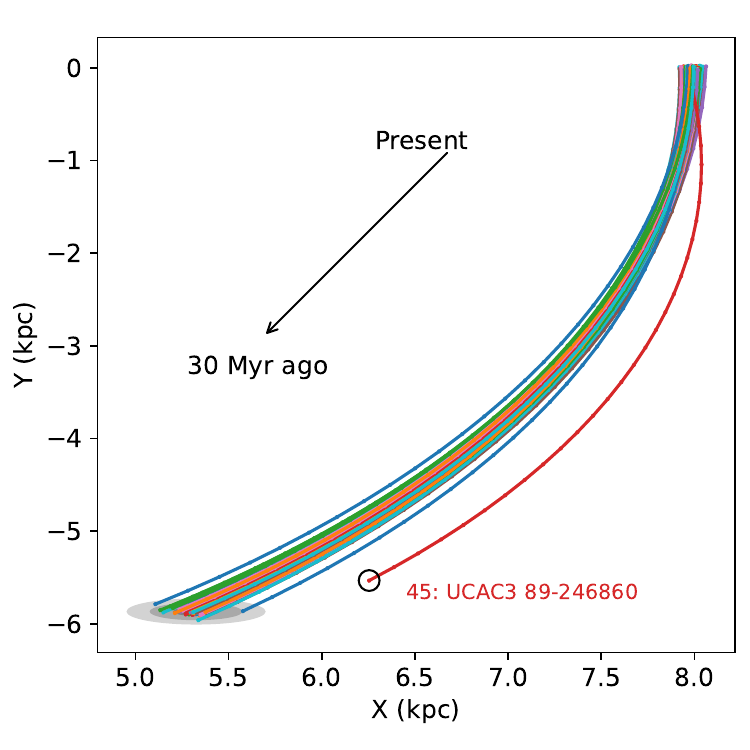}
   \caption{Stellar orbits were computed for all input members, tracing back from the present to 30 Myr ago using galpy \citep{bov15}. The resulting backtraced trajectories are depicted in the Galactocentric Cartesian coordinate system. Gray ellipsoids illustrate the 1 to 3-$\sigma$ of backtraced positions of the members 30 Myr ago. Circles annotate two kinematic outliers exceeding the 3-$\sigma$ threshold.}
    \label{fig:kin_outliers} 
\end{figure}

In the analysis of the kinematic age of the BPMG, using a reliable list of members is critical. We adopted the list of BPMG members from \cite{cou23}, comprising $N=76$ members divided into ``core samples'' ($N=25$) and ``extended samples" ($N=51$). We carefully examined and selected members based on six criteria for outlier evaluation: (1) known spectroscopic binaries from literature, (2) RV variability (RV variations greater than 2 \kms), (3) RV uncertainties larger than 1 \kms, (4) RUWE values from Gaia DR3 larger than 2, (5) inconsistent positions on the color-magnitude diagram (CMD) compared to other BPMG members (referred to as CMD outliers), and (6) kinematic outliers. Astrometric and photometric data are primarily sourced from Gaia DR3 \citep{gai16, gai23, bab23}, while RV data were obtained from \citet{cou23}, which compiled data from the literature (references therein). The assessment results for criteria (1), (2), and (4) are from \citet{cou23}.

Figure~\ref{fig:cmd_outliers} presents a color-magnitude diagram (CMD) of the entire input samples, with a theoretical isochrone model at 30 Myr displayed to assess the youth of the members. We used two isochrone models, PARSEC \citep{bre12} and BHAC15 \citep{bar15}, for spectral types of AFG (M$\geq$0.8M$_\odot$) and KM (M$<$0.8M$_\odot$), respectively. Faint colored circles represent other well-known NYMG members with different ages (TW Hydrae Association:$\sim$8 Myr, Tucana-Horologium Association:$\sim$30 Myr, and AB Doradus moving group:$\sim$100 Myr), as well as 500 randomly selected Gaia DR3 field stars within 100 pc. Most members align well between the sequences of TWA and AB Dor, indicating the intermediate age of BPMG and the well-prepared membership list. However, some members are fainter than expected for BPMG members of similar colors, suggesting that these stars may not be genuine BPMG members or that there are some peculiarities in their input data. 2MASS J04433761+0002051, StHA 182, and HD 207043 (star IDs: 34, 69, and 74 in Table~\ref{tab:members}) appear to be older than the other BPMG members. 

Kinematic outliers were assessed using backtraced stellar positions at 30 Myr in the past, approximately the upper age limit of the BPMG. Stellar positions were calculated by adopting the Milky Way's gravitational potential model (MWPotential2014) using galpy\footnote{http://github.com/jobovy/galpy \label{note1}} \citep{bov15}. Figure~\ref{fig:kin_outliers} displays the orbits of the entire input members on a Galactocentric Cartesian coordinate frame, tracing back from the present to 30 Myr ago. The Sun's present position is at (X, Y) = ($+8, 0$) kpc,  and nearly all NYMG members are concentrated around this position as expected. The Sun's position 30 Myr ago is at around (X, Y) = ($+5, -6$) kpc. 
Stars located outside the 3-$\sigma$ range at 30 Myr ago from the averages in the X, Y, or Z coordinates are defined as kinematic outliers. 
In Figure~\ref{fig:kin_outliers}, one star notably shows a deviating orbit from the other members. This star, UCAC3 89-246860 (star ID: 45), is classified as a kinematic outlier.
After removing these kinematic outliers, we additionally excluded 2MASS J18430597-4058047 (star ID: 61) due to its discrepant velocity toward the Galactic anti-center ($U=-0.75$ \kms) compared to the other adopted members ($U=-9.3\pm2.5$ \kms).

While not classified as outliers, one object was additionally excluded due to its lack of RV measurements. PSO J318.5338-22.8603 (star ID: 71) has only one RV measurement, and its RV variability cannot be evaluated.

\citet{cou23} used only ``core samples" ($N=25$) in their kinematic analysis of the BPMG, and 21 of these stars pass our outlier evaluation criteria. These stars are referred to as {\it core members} in this study. One reason for using our own evaluation of membership is to allow for the inclusion of more members from the ``extended samples". Including 14 members from the ``extended samples" of \citet{cou23} that pass our criteria, we generated a larger list called {\it inclusive members}. Table~\ref{tab:members} provides the entire input samples along with the results of our outlier evaluation. 
Their kinematic data are provided in Appendix~\ref{sec:appa}.  In this study, we primarily present the results obtained from {\it inclusive members}. 
The number of stars is 35 with mean and maximum RV uncertainties of 0.5 and 1.0 \kms, respectively.
Results based only on {\it core members} ($N=$21, with mean and maximum RV uncertainties of 0.5 and 1.0 \kms, respectively.), for easy comparison of the effects solely based on the chosen methods, are available in Table~\ref{tab:age_core} in Appendix~\ref{sec:appc}.

\startlongtable
\begin{deluxetable*}{ccccccc}   
 \tablecaption{Member Selection Results \label{tab:members}}
\tablehead{\colhead{Star ID\tablenotemark{a}} & \colhead{Designation} & \colhead{R.A.\tablenotemark{b}} & \colhead{Dec.\tablenotemark{b}} & \colhead{SpT} & \colhead{Membership status\tablenotemark{c}} & \colhead{Outlier category\tablenotemark{d}} \\
 & & \colhead{(hh:mm:ss.ss)} & \colhead{(hh:mm:ss.s)} & \colhead{Type} & \colhead{(1, 2, N)} & \colhead{(1$-$6)} 
}
\startdata
1  & RBS 38         & 00:17:23.82 & -66:45:12.6 & M2.5  & 1  & $-$ \\
2  & GJ 2006 A      & 00:27:50.37 & -32:33:07.1 & M3.5  & 1  & $-$\\
3  & HD 14082 B     & 02:17:24.84 & +28:44:29.3 & G2    & N  & 3 \\
4  & AG Tri A       & 02:27:29.35 & +30:58:23.4 & K8    & 1  & $-$\\
5  & CD-57 1054     & 05:00:47.20 & -57:15:24.3 & K8    & 1  & $-$\\
6  & V1841 Ori      & 05:00:49.30 & +15:26:59.8 & K2    & 1  & $-$\\
7  & BD-21 1074 A   & 05:06:49.97 & -21:35:09.4 & M1    & N  & 3 \\
8  & AO Men         & 06:18:28.18 & -72:02:40.3 & K3.5  & N  & 3 \\
9  & UCAC2 12510535 & 17:02:40.13 & -45:22:00.9 & M2    & 1  & $-$\\
10 & HD 164249 A    & 18:03:03.41 & -51:38:57.8 & F4.5  & 1  & $-$\\
11 & UCAC4 299-160704 & 18:04:16.18 & -30:18:29.0 & M4  & 1  & $-$\\
12 & 2MASS J18092970-5430532 & 18:09:29.72 & -54:30:55.0 & M4 & 1 & $-$\\
13 & 1RXS J184206.5-555426 & 18:42:06.98 & -55:54:26.8 & M3.5 & 1 & $-$\\
14 & Smethells 20   & 18:46:52.58 & -62:10:37.9 & M1   & 1 & $-$\\
15 & CD-31 16041    & 18:50:44.50 & -31:47:48.5 & K8   & N & 3 \\
16 & HD 181327      & 19:22:58.99 & -54:32:18.3 & F6   & 1 & $-$\\
17 & 1RXS J192338.2-460631 & 19:23:38.24 & -46:06:32.6 & M0 & 1 & $-$\\
18 & UCAC4 314-239934 & 19:48:16.55 & -27:20:32.8 & M2 & 1 & $-$\\
19 & TYC 7443-1102-1  & 19:56:04.41 & -32:07:38.8 & K9 & 1 & $-$\\
20 & UCAC4 284-205440 & 20:01:37.21 & -33:13:15.0 & M1 & 1 & $-$\\
21 & HD 191089        & 20:09:05.26 & -26:13:27.6 & F5 & 1 & $-$\\
22 & AU Mic           & 20:45:09.88 & -31:20:33.0 & M0 & 1 & $-$\\
23 & GSC 06354-00357  & 21:10:05.46 & -19:19:59.1 & M2 & 1 & $-$\\
24 & CPD-72 2713      & 22:42:49.26 & -71:42:22.0 & K7 & 1 & $-$\\
25 & WW PsA           & 22:44:58.19 & -33:15:03.7 & M4 & 1 & $-$\\
%\hline
26 & HD 203           & 00:06:50.20 & -23:06:27.9 & F3 & N & 2 \\
27 & LP 525-39        & 00:32:34.91 & +07:29:25.8 & M3 & N & 1  \\
28 & GJ 3076          & 01:11:25.63 & +15:26:19.9 & M5 & N & 1 \\
29 & Barta 161 12     & 01:35:14.03 & -07:12:52.2 & M4.3 & N & 1, 3 \\
30 & PM J01538-1459A  & 01:53:50.97 & -14:59:51.5 & M3 & N & 1, 3\\
31 & EPIC 211046195   & 03:35:02.15 & +23:42:34.4 & M8.5 & N & 3 \\
32 & StKM 1-433       & 03:57:33.97 & +24:45:09.9 & M2 & 2 & $-$ \\
33 & 51 Eri           & 04:37:36.18 & -02:28:25.8 & F0 & N & 3 \\
34 & 2MASS J04433761+0002051 & 04:43:37.67 & +00:02:03.4 & M9 & N & 5  \\
35 & V1005 Ori        & 04:59:34.88 & +01:46:59.2 & K8   & 2 & $-$ \\
36 & LP 476-207       & 05:01:58.83 & +09:58:57.1 & M4   & N & 1, 2, 3 \\
37 & AF Lep           & 05:27:04.78 & -11:54:04.3 & F7   & N & 2 \\
38 & UCAC4 287-007048 & 05:29:44.71 & -32:39:14.1 & M4.5 & 2   & $-$ \\
39 & V1311 Ori        & 05:32:04.51 & -03:05:30.0 & M2   & N & 1, 4 \\
40 & $\beta$ Pic      & 05:47:17.10 & -51:03:58.1 & A6   & N & 1, 3, 4 \\
41 & GSC 06513-00291  & 06:13:13.30 & -27:42:05.6 & M3.5 & N & 1, 3, 4 \\
42 & TWA 22 A         & 10:17:26.58 & -53:54:26.5 & M5   & N & 1, 4 \\
43 & $\alpha$ Cir     & 14:42:29.93 & -64:58:34.2 & A7   & N & 4  \\
44 & V343 Nor A       & 15:38:57.45 & -57:42:28.8 &G8   & N & 1, 3, 4 \\
45 & UCAC3 89-246860  & 16:12:05.11 & -45:56:26.0 & M4   & N & 2, 3, 6 \\
46 & TYC 8726-1327-1  & 16:57:20.24 & -53:43:33.0 & M3   & N & 1, 4 \\
47 & 2MASS J17020937-6734447 & 17:02:09.32 & -67:34:46.3 & M4  & 2 & $-$   \\
48 & 2MASS J17092947-5235197 & 17:09:29.45 & -52:35:20.9 & M3  & N & 3, 4  \\
49 & HD 155555 A      & 17:17:25.45 & -66:57:05.9 & G5   & N & 2, 3  \\
50 & CD-54 7336       & 17:29:55.07 & -54:15:49.7 & K1   & N & 1  \\
51 & HD 160305        & 17:41:49.03 & -50:43:29.1 & F9   & 2 & $-$  \\
52 & UCAC4 184-193200 & 17:44:42.57 & -53:15:48.5 & M3   & 2 & $-$ \\
53 & UCAC3 74-428746  & 17:48:33.74 & -53:06:12.8 & M2   & 2 & $-$  \\
54 & UCAC3 66-407600  & 18:05:54.92 & -57:04:31.9 & M2   & 2 & $-$  \\
55 & HD 165189        & 18:06:49.91 & -43:25:32.5 & A6   & N & 1, 3 \\
56 & V4046 Sgr        & 18:14:10.49 & -32:47:35.4 & K4   & N & 1, 2, 3 \\
57 & UCAC3 63-445434  & 18:16:12.41 & -58:44:07.9 & M3.5 & 2 & $-$  \\
58 & HD 168210        & 18:19:52.22 & -29:16:33.6 & G3   & N & 3  \\
59 & UCAC4 229-170667 & 18:28:16.52 & -44:21:48.6 & M2   & 2 & $-$  \\
60 & UCAC4 226-179372 & 18:28:35.25 & -44:57:28.9 & K7   & 2 & $-$   \\
61 & 2MASS J18430597-4058047 & 18:43:05.99 & -40:58:06.0 & M2 &  N & 6 \\
62 & HD 172555        & 18:45:26.98 & -64:52:18.9 & A7   & N & 2, 3  \\
63 & PZ Tel           & 18:53:05.90 & -50:10:51.3 & G9   & N & 1, 2, 3  \\
64 & TYC 6872-1011-1  & 18:58:04.17 & -29:53:05.4 & K8   & N & 1, 3, 4 \\
65 & 2MASS J19243494-3442392 & 19:24:34.99 & -34:42:40.6 & M4 & 2 & $-$  \\
66 & UCAC3 116-474938 & 19:56:02.98 & -32:07:19.8 & M4   & N & 1, 2, 3, 4   \\
67 & SCR J2010-2801   & 20:10:00.11 & -28:01:42.2 & M2.5 & N & 1, 3, 4   \\
68 & SCR J2033-2556   & 20:33:37.66 & -25:56:53.3 & M4.5& 2 & $-$  \\
69 & StHA 182         & 20:43:41.17 & -24:33:55.0 & M3.7 & N & 1, 3, 5  \\
70 & HD 199143        & 20:55:47.74 & -17:06:52.0 & F7   & N & 1, 2, 3, 4  \\
%71 & PSO J318.5338-22.8603 & 21:14:08.03 & -22:51:35.84 & L7 & N & 5  \\ 2MASS position at ep=2000
71 & PSO J318.5338-22.8603 & 21:14:08.18 & -22:51:38.2 & L7 & N & $-$\tablenotemark{e}  \\ % calculated @ ep=2016
72 & TYC 9114-1267-1  & 21:21:29.00 & -66:55:07.9 & K7   & N & 1, 2, 3, 4  \\
73 & 2E 4498          & 21:37:40.28 & +01:37:12.7 & M5   & N & 2, 3, 4  \\
74 & HD 207043        & 21:47:55.47 & -52:55:51.7 & G5   & N & 5  \\
75 & HD 213429        & 22:31:18.49 & -06:33:20.3 & F8   & N & 1, 2, 3  \\
76 & BD-13 6424       & 23:32:31.01 & -12:15:52.8 & M0   & 2 & $-$  \\   
\enddata
\tablecomments{Members are sourced from \citet{cou23}.}
\tablenotetext{a}{Star IDs are arranged in ascending order based on R.A., with {\it core members} listed first.}
\tablenotetext{b}{Positions are sourced from Gaia DR3, except for PSO J318.5338-22.8603, which is obtained from the 2MASS survey \citep{cut03}. All positions are at the 2016.0 epoch.}
\tablenotetext{c}{The membership statuses are coded as follows: (1) core member, (2) member but not core member, and (N) non-member or outlier.}
\tablenotetext{d}{The outlier categories are defined as follows: (1) known spectroscopic binary in literature, (2) RV variable, (3) RV uncertainties $>$1 \kms, (4) RUWE $>2$, (5) inconsistent CMD locations compared to other BPMG members, and (6) kinematic outlier.}
\tablenotetext{e}{This object has only a single RV measurement, making it impossible to evaluate its variability.}
\end{deluxetable*}

\subsection{Input Data}
Although the concept of kinematic age-dating analysis, finding the time when a group of stars occupies the smallest volume, is straightforward, implementing the method and selecting input data to demonstrate the concept is challenging. As illustrated before, a 1 \kms\ uncertainty in the input data causes about a 1 pc positional uncertainty over 1 Myr of traceback time. For typical NYMG members, Gaia data provide very precise astrometric information ($\lesssim20$ milli-arcsecond), and the resultant uncertainty in UVW caused by the typical Gaia uncertainty is smaller than 0.02 \kms. However, typical uncertainties in RV are an order of magnitude larger at least. Therefore, it is crucial to use the most precise input data and consider even a small systematic effect in the traceback age-dating. Furthermore, employing a method that systematically takes the uncertainty in input data into account is critical. \citet{cou23} demonstrated that stellar RV measurements can be systematically affected by two effects, gravitational redshift and convectional blueshift. These effects can alter RVs by $\sim0.6$ \kms. By applying the RV corrections, they concluded that the RV correction could contribute to $\sim+3$ Myr adjustment to the obtained kinematic age. However, the RV correction was applied only to the ``core samples" in their analysis. When the correction is applied to a larger population of candidate BPMG members, some additionally accepted BPMG members due to the RV correction can act in the opposite direction (i.e., lowering the kinematic age). We included the RV correction effect in our analysis as well, and our results with and without the RV correction are denoted as ``$\rm RV_{corr}$" and ``$\rm RV_{obs}$", respectively.

Large positional uncertainties in backtraced positions can result in a failed convergence at the birth time (e.g., \citealt{mam14}). Most of the previous kinematic age analyses on NYMGs did not incorporate the position uncertainties in the method. 
In our methods, we formally include the effect of input data uncertainties. Specifically, one of our three methods takes the positional uncertainties into account naturally by treating each star's position as a density function instead of treating it as a point source. For the other two methods we adopted, to assess the effects of uncertainties in backtraced positions, we employed a bootstrap resampling method of input data from the Gaussian distributions defined by uncertainties. We denote the results with formally assessed positional uncertainties as ``+Uncer", and cases without considering the effect of positional uncertainties are presented with ``$-$Uncer" denotation.

\section{Methods} \label{sec:methods}

In this study, we estimated the age of the BPMG through the application of three distinct methods: ``probabilistic volume calculation", ``mean pairwise distance calculation", and ``covariance matrix calculation".
As mentioned in Section~\ref{sec:memblist}, we computed the trajectories of stars using galpy \citep{bov15} by adopting the Milky Way's gravitational potential model, MWPotential2014. These stellar orbits were traced from their current positions back in time to 30 Myr ago.

\subsection{Method 1: Probabilistic Volume Calculation}

This method involves analyzing the spatial distribution of stars in an NYMG, considering measurement uncertainties, and calculating the volume they occupy.
In the context of studying an NYMG of stars spread over several tens of parsecs, individual stellar sizes are negligible, allowing us to treat each star as a point in 3D space. However, due to the presence of measurement uncertainties and their propagation in backtraced positions, we represent the true position of a star using a probability density function, which we call a position probability density function (PPDF). A PPDF takes the form of a 3D ellipsoid with axes determined by uncertainties in X, Y, and Z, where those uncertainties arise from uncertainties in the input parameters (R.A., Dec., and parallax).
When tracing back positions in time, the PPDF expands with time due to additional parameters involved (e.g., proper motion and RV) and their uncertainties. To calculate the volume occupied by $N$ stars while accounting for these positional uncertainties, we treat an NYMG as a collection of $N$ PPDFs. At each grid point in our coordinate system, we sum the values of these $N$ PPDFs to create a summed PPDF (sPPDF). The total star density of the NYMG can be obtained by summing sPPDF values across the entire grid domain, starting from group's center and extending infinitely. We define a volume to represent the space that contains 70\% of the total star density of the group. To calculate this enclosing volume, we employ the isosurface approach, using scikit-image.marching\_cubes \citep{scikitimg} to create an isosurface that encompasses 70\% of the total star density. The volume within this isosurface can be calculated using the vertices and faces of the 2D surface mesh. We refer to this method as ``Probabilistic Volume Calculation (PVC)".

As we have the relatively small number of members ($N=21$ for {\it core members} and $N=35$ for {\it inclusive members}), the sPPDF requires smoothing to reduce the influence of outliers. To achieve this, we applied data smoothing techniques to sPPDF, utilizing Kernel Density Estimation (KDE) in KDEpy \citep{odl18} with a Gaussian kernel. The bandwidth ($bw$) of the Gaussian kernel can be determined using Scott's rule-of-thumb \citep{sco92}, which suggests:   
\begin{math}
    bw = 1.06\sigma n^{-1/5}
\end{math}
where $\sigma$ represents the standard deviation of data and $N$ is the number of data. This rule is for data with a normal distribution. Because positions of NYMG members are not necessarily following a normal distribution and BPMG members show a larger dispersion along the X-axis, we decided to experiment with a range of $bw$ (7, 10, 15, 20 pc) instead of using a single value.

%Both two lists ({\it core members} and {\it inclusive members}) have a mean dispersion of $\sim$15 pc %[$\sigma_{XYZ}$=$(\sigma_X\sigma_Y\sigma_Z)^{1/3}$], yielding $bw\sim$7-8 pc. However, the data exhibit the %largest dispersion along the X-axis ($\sigma_X$), resulting in $bw$=16-17 pc. It is important to note that there %is no universally applicable bandwidth selection rule, and BPMG members do not necessarily follow a normal %distribution.
%Moreover, our primary objective in data smoothing is to eliminate localized (bumpy) features and obtain a more %representative volume for the group. We experimented with a range of bandwidths . 
To establish age upper and lower limits, we employed a leave-4-out jackknife resampling method, which includes all {\it core members} while randomly selecting 80\% of the remaining members ($N=14$). This results in 364 possible cases ($_{14}C_{11}$). 
The PVC method inherently calculates the case of $+$Uncer, and we only provide $+$Uncer calculations.
We also applied RV corrections for these two sets of calculations ($\rm RV_{corr}$ and $\rm RV_{obs}$).

\subsection{Method 2: Mean Pairwise Distance Calculation}

\begin{figure}[htb!]
    \plotone{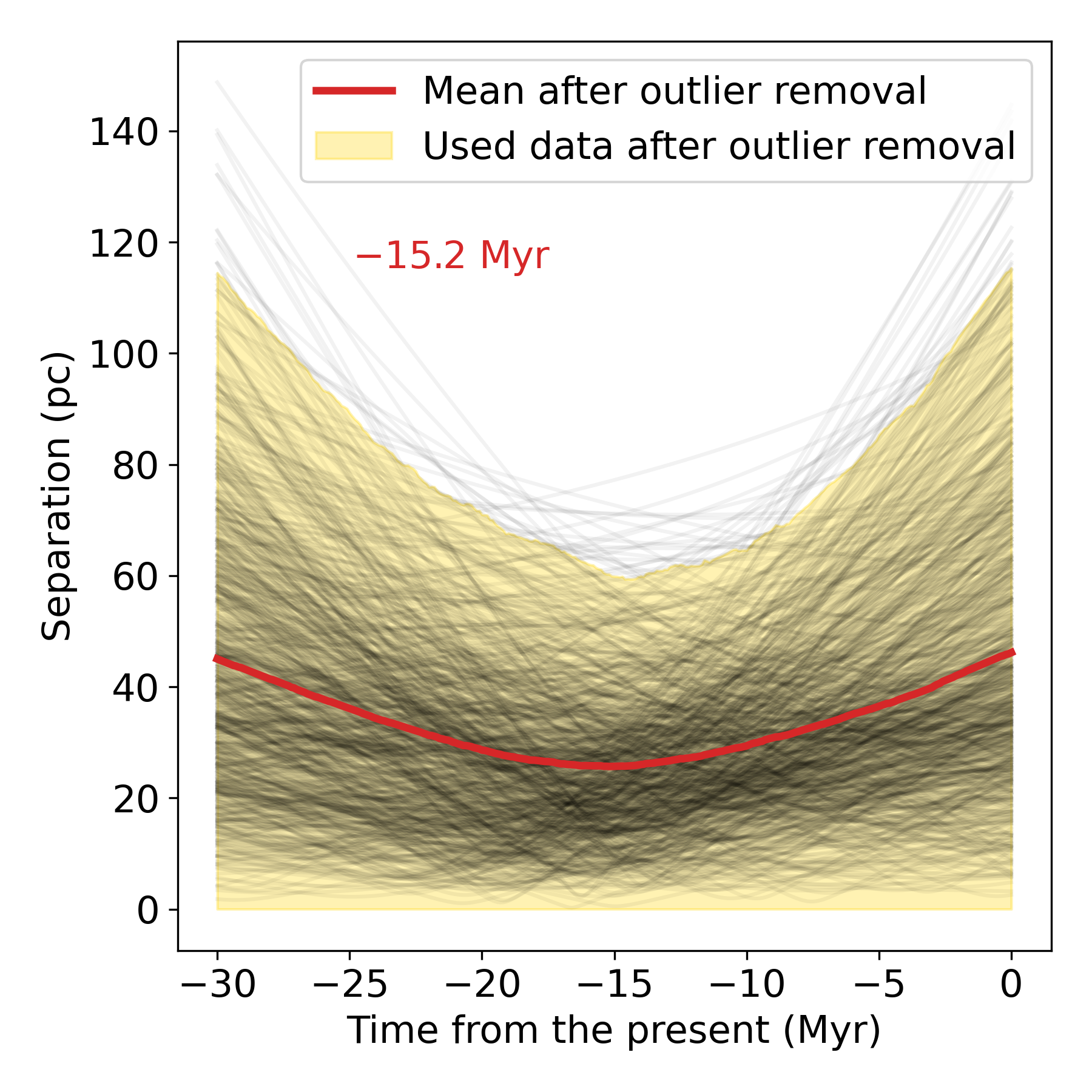}
    \caption{Distances between all pairs of {\it inclusive members} are depicted as gray lines over time. The mean distances, calculated by averaging after the removal of outliers\textsuperscript{\ref{note2}}, are shown as a red solid line. The time corresponding to the minimum distance is $-15.2$ Myr, written in the upper left corner.}
    \label{fig:dist_eg}
\end{figure}

In a dispersing NYMG, its members gradually move farther apart from each other over time. This means that the average distance in all pairwise distance combinations ($_{35}C_2$) decreases with traceback times until the group's birth time. Beyond the birth time, the average pairwise distance increases because of the imaginary expansion of the group before its existence. We attempted to find the time when the average pairwise distance is at its minimum. The approach employed in our calculation shares similarities with the one utilized in \citet{ker22}, although there are differences in details. In \citet{ker22}, they computed a median distance for each star to all other stars and determined the time at which the median distance was minimal. The minimal distance time (i.e., age) is obtained for all stars, and the final age is obtained from the distribution of ages. In our analysis, we obtained a time at which the mean separation of all pairs is minimum. From a single data set, we obtain an age instead of distribution of ages.
The number of available pairs for $N$ stars is
\begin{math}
  N(N-1)/2  
\end{math}, resulting in 595 cases.
Figure~\ref{fig:dist_eg} illustrates an example calculation with {\it inclusive members}. However, there are notably large separated pairs that can introduce a bias into the mean distance calculation. To mitigate this bias, we removed such outliers\footnote{Outliers are defined if the data is either $<$Q1-1.5*(Q3-Q1) or $>$Q1+1.5*(Q3-Q1), where Q1 and Q3 are 25th and 75th percentiles of the data, respectively.\label{note2}} before computing the mean distance. The yellow shaded area illustrates a region of data after excluding outliers, and the mean distance at each time is presented as a red solid line. The minimum mean pairwise distance occurs at $-15.2$ Myr in this example.

To estimate the age uncertainty, we used bootstrap resampling, performing the calculation 10,000 times.
This calculation is performed considering data uncertainties (+Uncer and $-$Uncer) and RV corrections ($\rm RV_{corr}$ and $\rm RV_{obs}$).

\subsection{Method 3: Covariance Matrix Calculation}

The positions of stars in 3D space can be fitted as an ellipsoid, whose volume can be represented as a covariance matrix. We computed the geometric mean of the standard deviation along the principal axes (i.e., determinant of the covariance matrix) at each time. Subsequently, we found the time at which this mean size reaches its minimum (a similar calculation was performed in the study by \citealt{mir20}).

In addition to the above, akin to our approach for pairwise distance calculations, we evaluated ages in cases of considering RV correction and data uncertainties and employed the bootstrap resampling technique.

\section{Results} \label{sec:results}

\subsection{Method 1: Probabilistic Volume Calculation (PVC)} 

\begin{figure}[htb!]
    \plotone{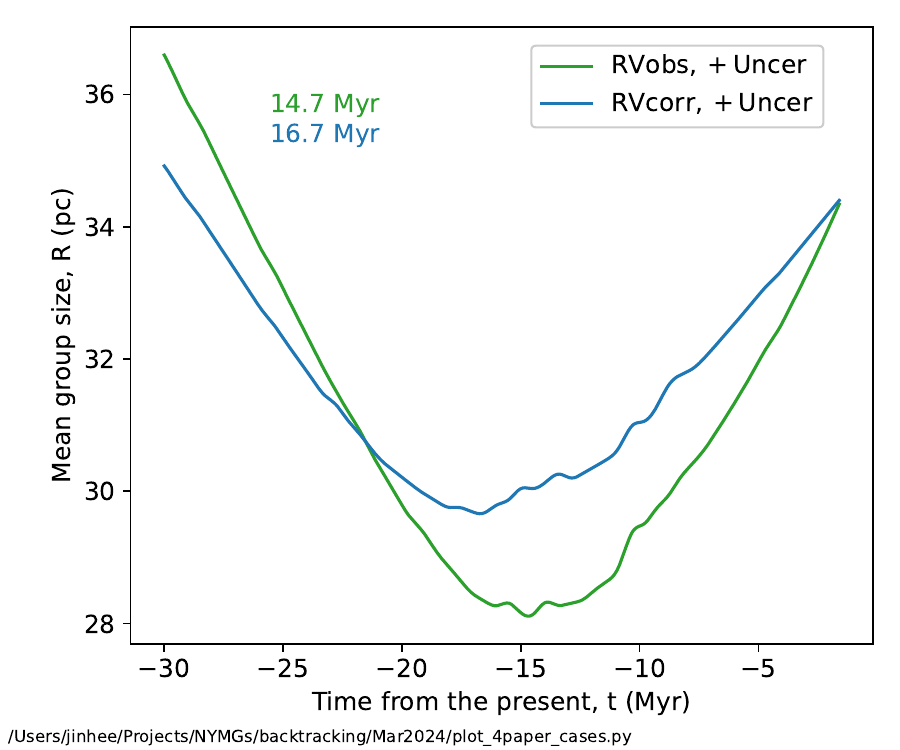}
    \caption{An example calculation of group size ($\sim$volume$^{1/3}$) with a bandwidth ($bw$) of 10. The group size is estimated in  two different scenarios, considering data uncertainties and RV correction. The volume is defined by the isosurface enclosing 70\% of the total star density. The times corresponding to the minimum volume for each scenario are denoted in the upper left corner of the figure.}
\label{fig:vol_eg}    
\end{figure}

\begin{figure}[htb!]
    \includegraphics[width=0.99\linewidth]{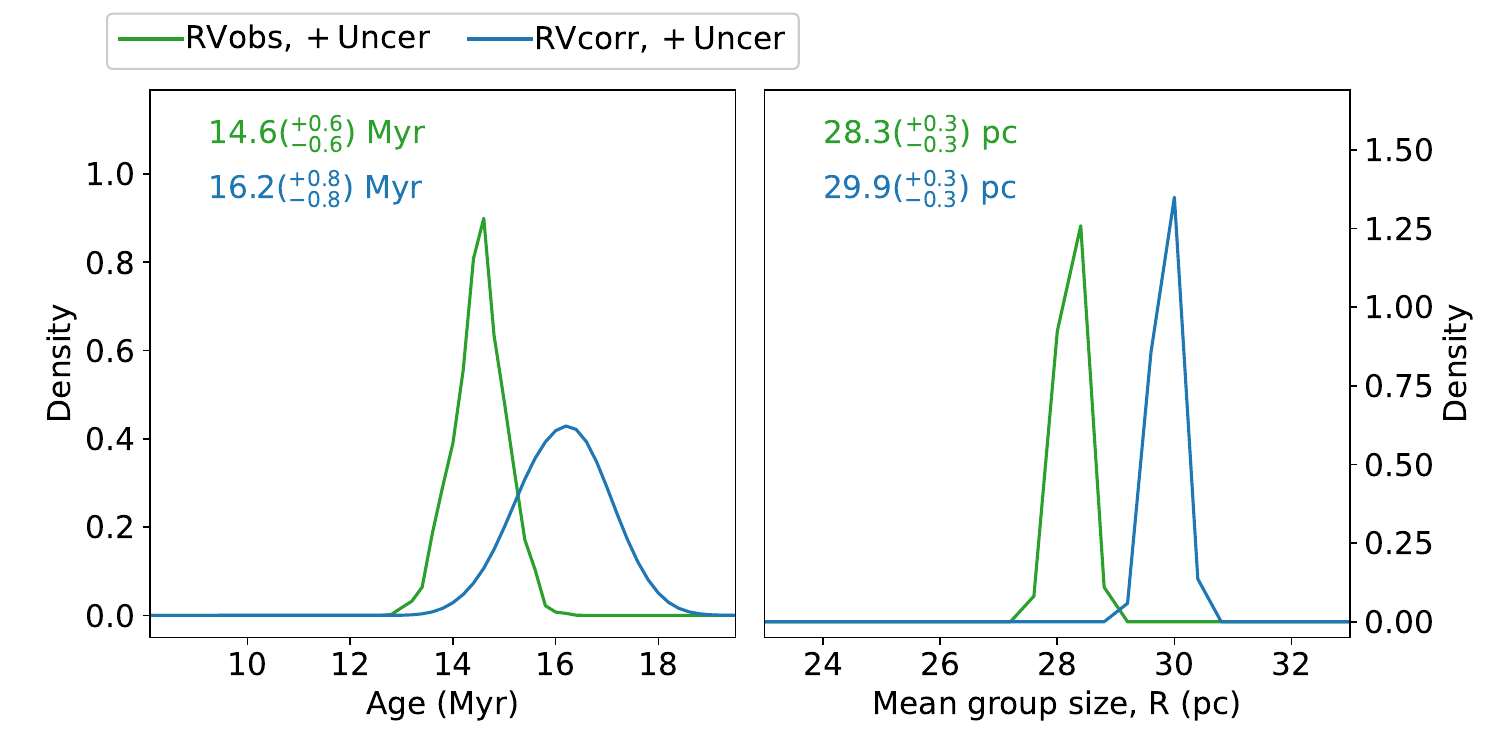}
    \caption{Distributions of age (left) and mean size (right) from PVC with $bw=10$. These results are obtained through jackknife resampling, which includes all {\it core members} and 80\% randomly selected remaining {\it inclusive members} in each resampling.  The medians with 68\% confidence intervals are denoted in the upper of the figure.}
\label{fig:vol}    
\end{figure}

The volume ($V$) obtained from the PVC method is translated into the mean group size ($R$) with a simplifying assumption ($V=4/3\pi R^3$). 
 Figure~\ref{fig:vol_eg} illustrates the mean size of the group as a function of time. At the present time ($t=0$), the mean size of the group is $\sim34$ pc. As we trace back in time, the mean group size ($R$) decreases to reach a minimum and subsequently increases for further back in time. We conducted these calculations for the mean group size as a function of time in two different scenarios, taking into account the effects of RV correction. The age from each scenario was determined by finding the time of the minimum mean group size. In Figure~\ref{fig:vol_eg}, the ages are 14.7 and 16.7 Myr for case of $\rm RV_{obs}$+Uncer and $\rm RV_{corr}+$Uncer, respectively. These calculations were carried out using jackknife resampling. The 364 calculations are performed for each scenario.
The left panel of Figure~\ref{fig:vol} displays distribution of the ages, while the right panel shows the mean group size $R$. The median and 68\% confidence intervals are indicated as the age (left panel) and the mean group size (right panel) with uncertainties. Table~\ref{tab:age} summarizes the results under various scenarios, including different bandwidths for KDE.

When we analyzed data without considering RV correction [$\rm RV_{obs}, +$Uncer], the estimated ages are 14$-$15 Myr with all $bw$ values. When RV correction is applied [$\rm RV_{corr}, +$Uncer], the estimated ages increase by $1-2$ Myr, resulting in an age of approximately $16-17$ Myr.
These trends are consistent with all $bw$ values, but larger $bw$ values tend to yield slightly younger age estimates. Nonetheless, the differences in age estimates between different $bw$ values are generally less than 1 Myr.

The mean group size at the time of minimum volume (i.e., its birth time) is approximately 30 pc as seen in Figures~\ref{fig:vol_eg} and~\ref{fig:vol} (right). The mean sizes resulting from full calculations with resampling and different $bw$ values are provided in Appendix~\ref{sec:appb}. These sizes vary from 25 to 45 pc, with an increase in the $bw$ values. More smoothing of the density cube with a larger $bw$ results in larger group size estimates, as expected.

\begin{deluxetable}{lcc}
   \tablecaption{Ages Derived Through Three Distinct Methods \label{tab:age}} 
   \tablehead{ & \colhead{$-$Uncer} & \colhead{$+$Uncer} \\
       & \colhead{(Myr)} & \colhead{(Myr)}}
       \startdata
       \multicolumn{3}{c}{Method 1: Probabilistic Volume Calculation (PVC)} \\
       \hline
       \multicolumn{3}{c}{$bw$=7} \\
  %     \cmidrule(ll){3-3}
       $\rm RV_{obs}$ & $-$ & 14.8$(^{+0.5}_{-0.5})$\\
       $\rm RV_{corr}$ & $-$ & 16.7$(^{+0.9}_{-0.9})$  \\
       \multicolumn{3}{c}{$bw$=10} \\
 %      \cmidrule(ll){2-3}
       $\rm RV_{obs}$ & $-$ & 14.6$(^{+0.6}_{-0.6})$\\
       $\rm RV_{corr}$ & $-$ & 16.2$(^{+0.8}_{-0.8})$ \\
       \multicolumn{3}{c}{$bw$=15} \\
  %    \cline{3-3}
       $\rm RV_{obs}$ & $-$ & 14.4$(^{+0.6}_{-0.6})$ \\
       $\rm RV_{corr}$ & $-$ & 15.9$(^{+0.7}_{-0.7})$ \\
       \multicolumn{3}{c}{$bw$=20} \\
  %     \cline{2-2}
       $\rm RV_{obs}$ & $-$ & 14.2$(^{+0.6}_{-0.6})$ \\
       $\rm RV_{corr}$ & $-$ & 15.6$(^{+0.8}_{-0.8})$ \\
        \hline       
        \multicolumn{3}{c}{Method 2: Mean Pairwise Distance Calculation} \\
        \hline
       $\rm RV_{obs}$ & 15.6$(^{+0.9}_{-1.0})$ & 14.5$(^{+1.3}_{-1.4})$ \\
       $\rm RV_{corr}$ & 17.8$(^{+1.2}_{-1.5})$ & 15.9$(^{+1.9}_{-2.2})$ \\
       \hline
       \multicolumn{3}{c}{Method 3: Covariance Matrix Calculation} \\
       \hline
       $\rm RV_{obs}$  & 16.7$(^{+0.9}_{-0.8})$ & 15.6$(^{+1.4}_{-1.3})$\\
       $\rm RV_{corr}$ & 20.5$(^{+3.1}_{-1.5})$ & 18.3$(^{+2.9}_{-2.1})$ \\              
       \hline       
        \multicolumn{3}{c}{Summary\tablenotemark{a}} \\
       \hline
       $\rm RV_{obs}$  & 16.2$\pm$0.7 & 14.7$\pm$0.5 \\
       $\rm RV_{corr}$  & 18.3$\pm$1.4 & 16.3$\pm$0.7 \\       
\enddata
   \tablecomments{The results are obtained using the {\it inclusive} members. The reported uncertainties correspond to the 68\% confidence interval.}       
   \tablenotetext{a}{The summary ages were determined through a weighted mean of three methods. We utilized an age with $bw$=10 for the PVC, and we adopted larger uncertainties.}       
\end{deluxetable}

\subsection{Method 2: Mean Pairwise Distance Calculation}
\label{sec:method2}

\begin{figure}[htb!]
   \includegraphics[width=0.99\linewidth]{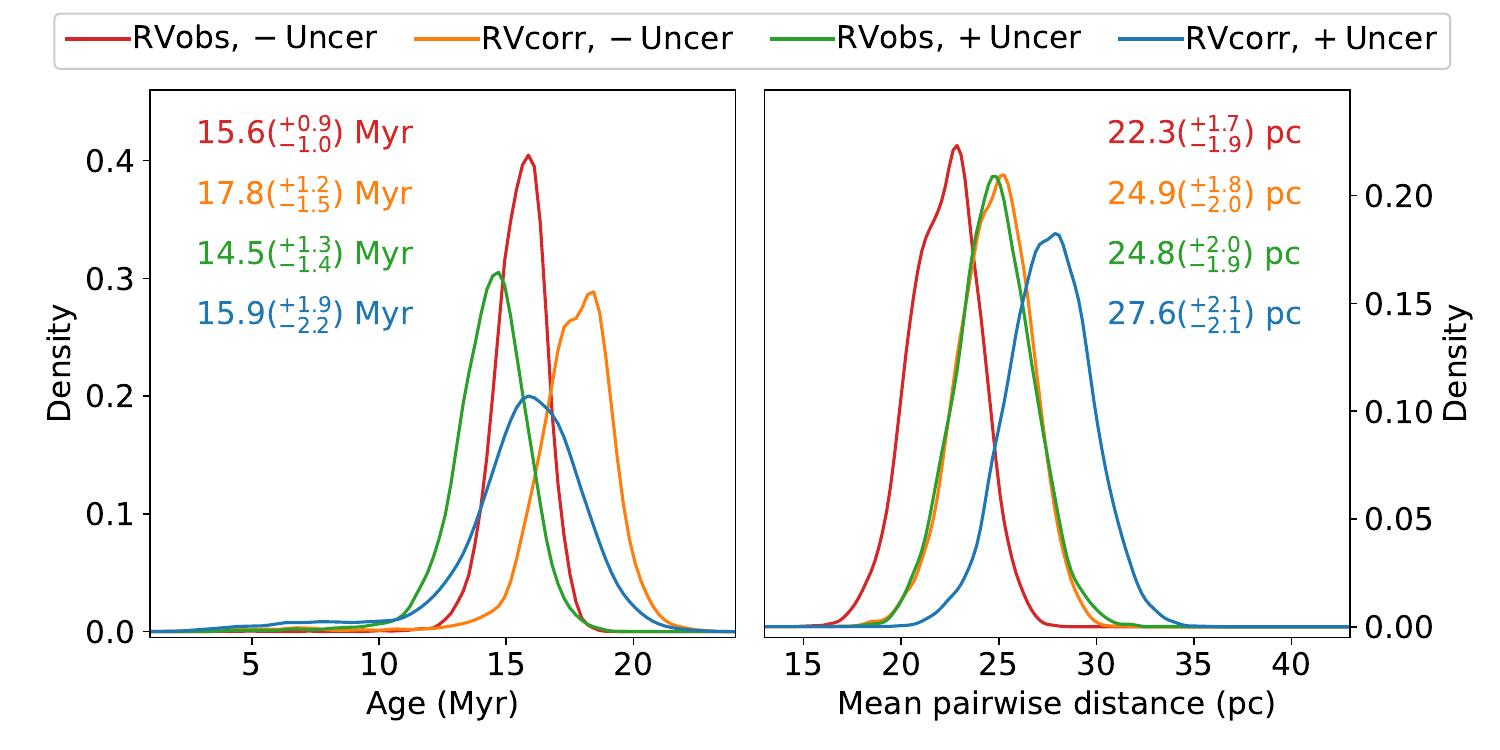}
    \caption{Distribution of age (left) and mean pairwise distance (right) from Method 2. These results are obtained through 10,000 rounds of bootstrap resampling. The medians with 68\% confidence intervals are denoted in the upper of the figure.}
    \label{fig:dist}
\end{figure}

Mean pairwise distances between all pairs of stars were calculated as a function of time. For a given dataset, the time of the minimum mean distance was found and considered as the birth time of the group. The calculations were performed through 10,000 rounds of bootstrap resampling. The results of calculations are presented in Figure~\ref{fig:dist}. The application of RV correction ($\rm RV_{corr}$) tends to increase the estimated age, while considering data uncertainties (+Uncer) has the opposite effect, resulting in a decrease in the calculated age. 

The calculated ages for the four scenarios are listed in Table~\ref{tab:age}.
These findings align closely with the case using a $bw$ of 15 in the PVC analysis. Furthermore, the estimated mean pairwise distances are approximately $22-28$ pc (presented in Appendix~\ref{sec:appb}).

\subsection{Method 3: Covariance Matrix Calculation}
\label{sec:method3}

\begin{figure}[htb!]
    \centering
    \includegraphics[width=0.99\linewidth]{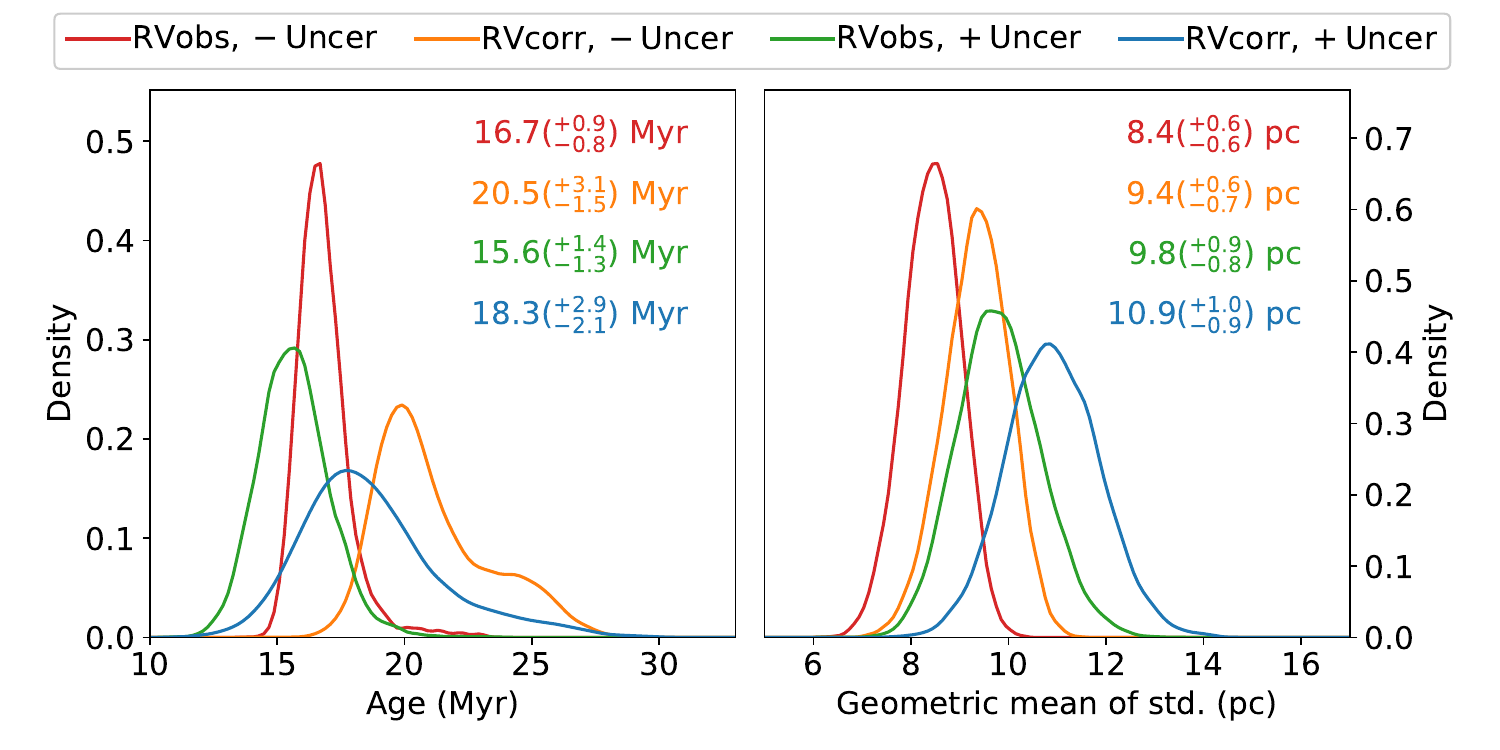}
    \caption{Distributions of age (left) and determinant of covariance matrix (right). These results are obtained through 10,000 rounds of bootstrap resampling. The medians with 68\% confidence intervals are denoted in the upper of the figure.}
    \label{fig:covmat}
\end{figure}

Similar to the previous analyses conducted using the two methods described earlier, we computed the covariance matrix considering data uncertainties and RV correction. The time of the minimum size of the position ellipsoid was considered as the birth time of the group hence the age of the group, and age upper and lower limits were determined through bootstrap resampling. The results are depicted in Figure~\ref{fig:covmat} and summarized in Table~\ref{tab:age}.

Consistent with the outcomes of the two preceding methods, we observe that the application of RV correction tends to increase the estimated ages, whereas considering data uncertainties leads to a reduction in the resultant ages. 

The lengths of the semi-major axes of the ellipsoid correspond to the standard deviation (1-$\sigma$), which would encompass approximately $\sim30\%$ of the group ($0.68^3$) assuming a normal distribution of members. The estimated mean size obtained from the covariance matrix is approximately 10 pc (provided in Appendix~\ref{sec:appb}), and it notably appears to be smaller than the sizes obtained through the PVC method, primarily due to the size estimation differences. However, when mean sizes are calculated using 2-$\sigma$ ($0.95^3\sim0.86$), the range becomes 20 pc, aligning well with the results obtained from the PVC method with $bw=7$.

\section{Discussion} \label{sec:discussion}

Kinematic analysis serves as a semi-fundamental method for age dating a stellar group, offering the advantage of being insensitive to chosen models \citep{sod10}. However, this method is sensitive to the composition of the member list, and the presence of outliers can significantly bias age estimations \citep{mir20}. 
The member lists utilized in this study were meticulously generated, with discrepant members excluded as explained in Section~\ref{sec:memblist}. The calculations presented here were performed using {\it inclusive members} ($N$=35), and the calculations derived from {\it core members} ($N$=21) provide consistent results (provided in Appendix~\ref{sec:appc}). The differences in age estimations obtained with {\it inclusive members} and {\it core members} are typically less than 1 Myr.

The choice of the size estimation metric plays a role in kinematic age estimation.
In this study, three methods were employed, and they consistently yield similar results, as summarized in Table~\ref{tab:age}. 
Ages obtained from the PVC and mean pairwise distance calculation exhibit similarities, whereas ages derived from the covariance matrix calculation tend to be approximately $1-2$ Myr older on average.

The major factors that influence age estimations are data uncertainties and RV correction. When data uncertainties and RV corrections are not considered [$\rm RV_{obs}, -$Uncer], the age estimates range from 16 to 17 Myr. The inclusion of RV correction to the data [$\rm RV_{corr},-$Uncer] resulted in an increase in the age estimation to $18-21$ Myr as reported in \citet{cou23}. Conversely, when we consider data uncertainties [$\rm RV_{obs}$,+Uncer], the age estimates decreased to $\rm 14-16$ Myr. Finally, when both data uncertainties and RV correction are taken into account [$\rm RV_{corr}$,+Uncer], the age estimation falls within the range of $\rm 16-18$ Myr. 
\citet{mir18} and \citet{cou23} have demonstrated that taking input data uncertainties into the calculation always decreases the kinematic age estimates. The primary reason for such a behavior could be due to the fact that the kinematic age dating is a calculation with a half-open boundary whose anchor point is the values defined by the present-day positions and uncertainties. The volume of a group members at present is well restricted with small uncertainties while the volume is more uncertain in the past because of the effect of the positional uncertainty propagation. 
When one adds errors in simulated input data sets as in the case of \citet{mir18}'s simulation, the minimum volume obtained by the calculation becomes larger than the true value.
Since the volume is anchored at the present, the only way the curve of the volume as a function of traceback time can move with added uncertainties is to the direction of the upper-right corner in Figure 3 of \citet{mir18}. As a result, a simulation with added uncertainties in the input data ends up with a younger minimum volume time and a larger minimum volume always. The magnitude of this effect should be dependent on the relative magnitudes of the intrinsic velocity dispersion at birth and observational uncertainties. As the intrinsic dispersion is unknown, predicting the contribution of observational uncertainties to the decrease in kinematic age is challenging. Therefore, the younger ages observed in the $+$Uncer cases ($\sim$2 Myr younger) in this study cannot solely be attributed to the effect of the half-open boundary problem.

The final reported age of BPMG was determined from the weighted mean of all age estimates considered in this paper.
We obtained a weighted mean age from the three methods in Table~\ref{tab:age} for the [$\rm RV_{corr}$+Uncer] case. 
The upper and lower age limits were manually selected to encompass the age ranges of the four ages listed in the summary panel of Table~\ref{tab:age}, as the resultant age uncertainty from the weighted mean is too small ($<$1 Myr) and we cannot tell if any one method out of three produces more reliable result than other methods. 
 Our final BPMG kinematic age reported in this paper is 16.3$^{+3.4}_{-2.1}$ Myr.

\subsection{Comparison with Other Studies on the Kinematic Age of BPMG} 

In addition to the three methods applied in this study, several different approaches have been employed in previous studies to determine the kinematic ages of the BPMG.  
\citet{ort02} determined the expansion age by calculating the maximum distance between members. \citet{son03} observed the time when the BPMG members were most compact in the XY space. The standard deviation of members along each axis ($\sigma_X$, $\sigma_Y$, and $\sigma_Z$) was used in \cite{mam14}.  They concluded that the traceback analysis is not applicable to the BPMG with a large range of the kinematic age ($13-58$ Myr). 
 \citet{cru19} assumed a group model at the group's birth time in 6D space and evaluated the evolution of the model along the Galactic potential, then determined the kinematic age by comparing it to the observation data. The covariance matrix, median absolute deviation, and minimum spanning tree were used in studies of \citet{mir18, mir20} and \citet{cou23}.

The differences in age estimations between these earlier studies and our present study can be attributed to variations in the group size estimation metric, the composition of member lists, and input data. 
Our age estimate ($16.3^{+3.4}_{-2.1}$ Myr) falls within a range between the estimates from earlier studies (\citealt{son03}; 12 Myr, \citealt{ort02,ort04}; $11-12$ Myr), and more recent studies (\citealt{cou23}; 20.4 Myr, \citealt{mir20}; 18.5 Myr, \citealt{cru19}; 17.8 Myr).

\subsection{Comparison with Ages Determined by Other Methods}

The age of BPMG has been estimated using various methods, including isochrone fitting, lithium depletion boundary (LDB), and empirical approaches based on signs of activities. 

\cite{zuc01} reported seventeen members and estimated the age as 12 Myr based on the H-R diagram, H-$\alpha$ emission, X-ray counts from ROSAT, and equivalent width of lithium.  During the same period, \cite{bar99} determined the age of group as 20$\pm$10 Myr through isochrone fitting.

In the analysis of age of the BPMG using kinematic ages, \cite{mam14} also incorporated isochronal analyses. They claimed that the A-type members are all on the zero-age main sequence, implying an age of greater than 20 Myr. They also noted that late-F and G pre-main sequence stars align with the isochronal age of 22 Myr.

\begin{figure}[htb!]
    \centering
    \includegraphics[width=0.99\linewidth]{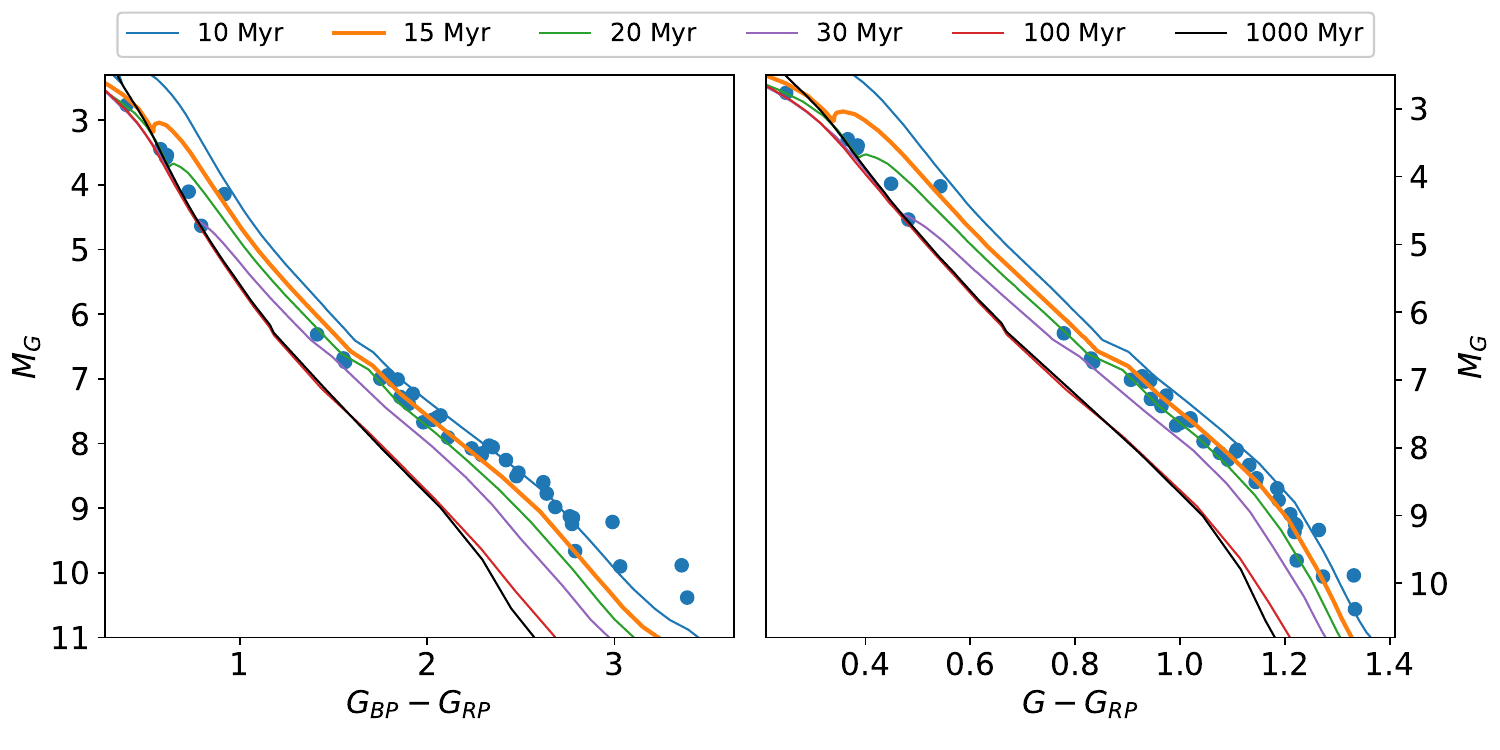}
    \caption{Color-magnitude diagrams of the BPMG members compared to the theoretical isochrones (PARSEC \citep{bre12} and BHAC15 \citep{bar15} for $M\geq 0.8M_\odot$ and $M< 0.8M_\odot$, respectively).}
    \label{fig:cmd_iso}
\end{figure}

\cite{bel15} determined the isochronal age of the BPMG using a CMD ($V-J$ vs $M_V$) and four isochronal models. They concluded that the best-matching age for the BPMG with 97 members is $24\pm3$ Myr.
Recently, \cite{ujj20} assessed the age of the BPMG using photometric data from Gaia DR2, resulting in an estimated age of 19.38 Myr. 
In a study of \cite{lee22}, the age of the BPMG was evaluated using combined isochrones that incorporated the PARSEC model \citep{bre12} for massive stars ($\geq 0.7 M_\odot$) and the BHAC15 model \citep{bar15} for low-mass stars ($< 0.7 M_\odot$). The age estimation was approximately 10 Myr, primarily influenced by the low-mass members for which age estimations were from the BHAC15 model.
Figure~\ref{fig:cmd_iso} presents color-magnitude diagrams of the {\it inclusive members} in comparison to the theoretical isochrone models \citep{bre12, bar15}. Most of the members are in good agreement with the models within the range of 15 to 20 Myr, while a few members align with the 10 Myr or 30 Myr models.

As a stellar group ages, the fraction of stars with gaseous or dusty disks decreases. The BPMG is recognized as a disk-rich group with a large debris disk fraction ($>37\%$; \citealt{reb08}, $57-75\%$; \citealt{naj22}). \citet{naj22} shows that the dust luminosity of the BPMG is similar to those of stars with ages of $10-40$ Myr.

The lithium depletion boundary (LDB) method has been actively employed to estimate the age of stellar groups including BPMG, as the method is believed to be based on better constrained stellar physics than CMD isochrone fittings where stellar luminosities can oftentimes be more affected by various effects such as episodic accretion history, stellar spots, etc.
Several studies assessed the LDB age of the BPMG, ranging from approximately 20 to 25 Myr \citep{bin14, mal14, mes16}. 
Notably, \cite{mes16} considered the impact of activity by correcting Li and employing models with magnetic fields, resulting in an age estimate of 25$\pm$3 Myr.  
However, it is essential to note that the LDB age determination is typically applicable within a range of 20 to 200 Myr, despite its relatively high precision of $\pm10\%$ \citep{bur04}. 
The LDB age of the BPMG falls close to the lower limit of LDB method's applicability, suggesting that the LDB age may be subject to greater uncertainty.

\subsection{Origin of the BPMG}
\begin{figure*}[htb!]
    \includegraphics[width=0.99\linewidth]{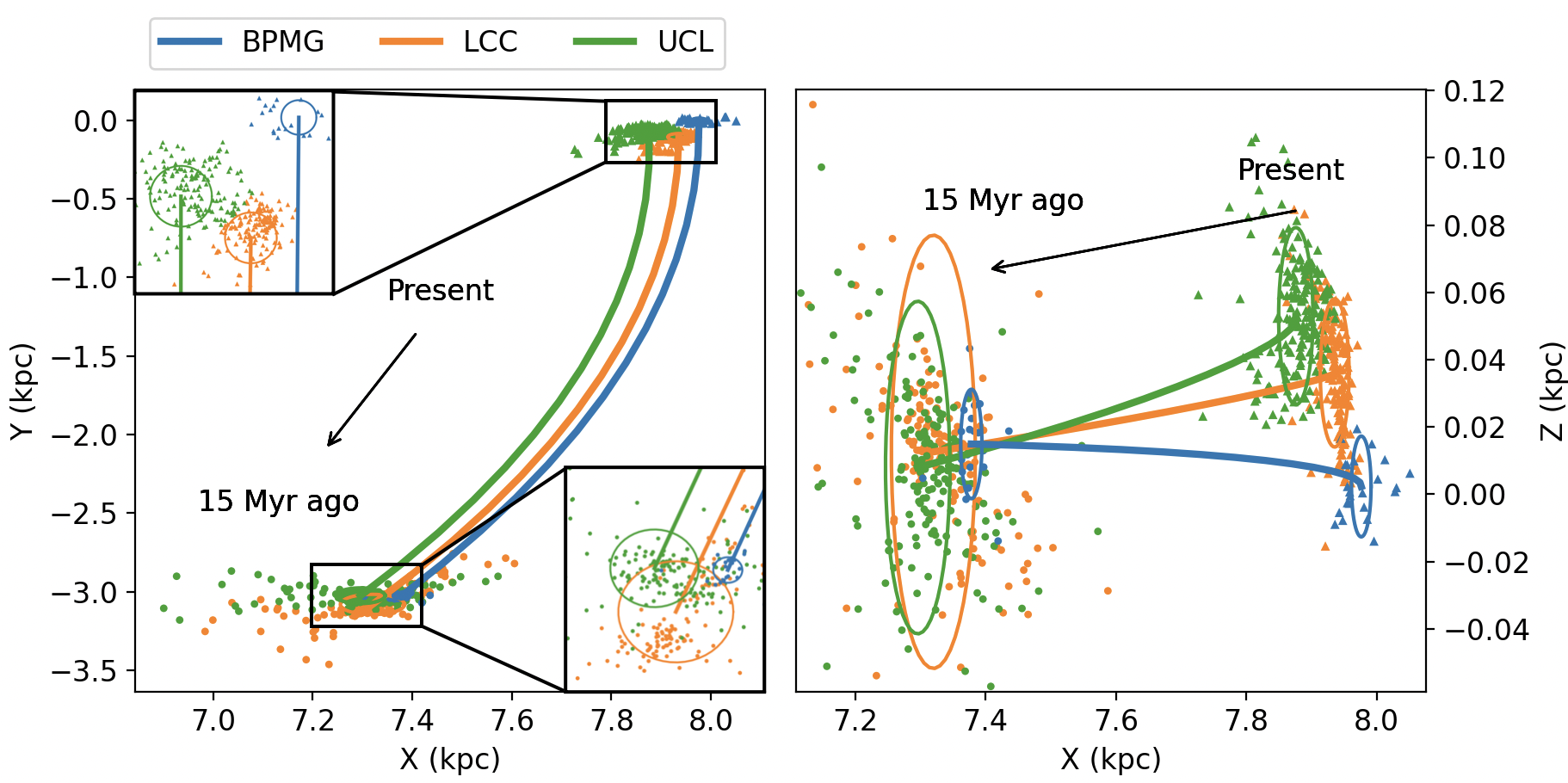}
    \caption{Traceback positions of BPMG members and two subgroups (LCC and UCL) of the Sco-Cen complex. The members of Sco-Cen are taken from \citet{zee99}, \citet{mam02}, and \citet{son12}. The mean traceback positions for each group are presented as thick solid lines. The time integration is taken from the present to $-15$ Myr. For better presentation, the positions of members are presented only at the time of the present (triangles) and the $-15$ Myr (dots).}
    \label{fig:bpmg_scocen}
\end{figure*}

The origin of the BPMG has been hypothesized to result from interaction with the nearby OB association, the Scorpius Centaurus complex (Sco-Cen). \cite{ort02, ort04} analyzed the relationship between the BPMG and two of these subgroups, Lower Centaurus Crux (LCC) and Upper Centaurus Lupus (UCL). 
Figure~\ref{fig:bpmg_scocen} illustrates the traceback orbits of these groups from the present to 15 Myr ago. The members of Sco-Cen are taken from \citet{zee99}, \citet{mam02}, and \citet{son12}. Astrometric data, including RV, are taken from Gaia DR3 \citep{gai23}. This improved quality of data provides a more accurate relationship between these three groups. 
In Figure~\ref{fig:bpmg_scocen}, the group sizes are represented as circles with radii determined by the geometric mean of the standard deviation (1-$\sigma$) of the member distribution. The groups are well separated at present, while they largely overlapped 15 Myr ago, at the time of around these groups' formation.
This supports the idea that the formation of these three groups might be closely related, as argued in \citet{ort02, ort04}.

\citet{zuc22} provided a comprehensive picture of the formation and evolution history of young nearby stars. They explained that LCC and UCL were formed at $\sim15-16$ Myr ago at the center of the local bubble due to the explosion of $N\sim15$ supernovae. The age of the BPMG, around $16$ Myr, naturally supports the idea that the origin of BPMG is closely related to the formation of the Sco-Cen complex. In the outskirt region, with relatively lower-mass members and lower numbers, the BPMG might have formed at around the time of the formation of the local bubble. 
This scenario is consistent with the findings of \citet{son12}, which suggest that the BPMG is slightly older than LCC and UCL according to the LDB analysis.

\section{Conclusions} \label{sec:conclusion}

As a benchmark NYMG, the age of the BPMG holds significant importance in shaping our understanding of young stars and the astrophysical properties of planets, as well as the evolution of loose stellar associations.  

In this study, we estimated the kinematic age of the BPMG using three distinct methods: probabilistic volume calculation, mean pairwise distance calculation, and covariance matrix calculation. Notably, we carefully accounted for the effects of data uncertainties. We also considered RV corrections due to convectional blueshift and gravitational redshift, as demonstrated in \citet{cou23}. 

All three methods employed in our study yielded consistent age estimations, falling within the range of $14-20$ Myr. RV correction was observed to increase the age estimates by approximately $2-4$ Myr, while data uncertainties had the opposite effect, reducing age estimations by $1-2$ Myr. When accounting for all these influences [$\rm RV_{corr}$,+Uncer], we arrived at an age estimation of $16.3^{+3.4}_{-2.1}$ Myr. 

While this age estimate falls within the known BPMG age range, which spans from 10 to 25 Myr, it is notably younger than the recently accepted age of $\sim20-25$ Myr. This relatively young age estimation underscores the necessity for a revisitation of systematic analyses of the BPMG's age through other age estimation methods and recalibration to achieve consistent age estimations.

%% Also note that the akcnowlodgment environment does not support long amounts of text. If you have a lot of people and institutions to acknowledge, do not use this command. Instead, create a new \section{Acknowledgments}.

\section{Acknowledgments}
We acknowledge Jae-Joon Lee and the anonymous referee for discussion and comments that greatly improved the this paper. 
J. L. acknowledge the support by the Korea Astronomy and Space Science Institute (KASI) grant funded by the Korean government (MSIT; No. 2024-1-860-02, International Optical Observatory Project).

This work has made use of data from the European Space Agency (ESA) mission
{\it Gaia} (\url{https://www.cosmos.esa.int/gaia}), processed by the {\it Gaia}
Data Processing and Analysis Consortium (DPAC,
\url{https://www.cosmos.esa.int/web/gaia/dpac/consortium}). 


% KASI Cloud
The authors acknowledge that the computational work reported on in this paper was (partly) performed on the KASI Science Cloud platform supported by Korea Astronomy and Space Science Institute.

\vspace{5mm}
%\facilities{HST(STIS), Swift(XRT and UVOT), AAVSO, CTIO:1.3m,


\software{  
NumPy \citep{numpy20}, 
SciPy \citep{scipy20}, 
Matplotlib \citep{matplotlib07}, 
KDEpy \citep{odl18}, 
scikit-image \citep{scikitimg}
          }



\clearpage

\appendix

\section{Input data \label{sec:appa}}

%\startlongtable
%\begin{rotatetable}
\begin{deluxetable*}{ccclllllllllll}
\tabletypesize{\scriptsize}
\rotate
    \tablecaption{Kinematic Data of the Full Input Samples \label{tab:fulldata}}
        \tablehead{\colhead{Star ID} & \colhead{Designation} & \colhead{SpT} & \colhead{R.A.}  & \colhead{Dec.} & \colhead{Plx} & \colhead{eplx}  & \colhead{pmRA} & \colhead{epmRA} & \colhead{pmDE} & \colhead{epmDE} & \colhead{RV} & \colhead{eRV} & \colhead{RUWE} \\ 
    &   & & \colhead{(deg)} & \colhead{(deg)}  & \colhead{(mas)} & \colhead{(mas)} & \colhead{(\masyr)} & \colhead{(\masyr)} & \colhead{(\masyr)} & \colhead{(\masyr)} & \colhead{(\kms)} & \colhead{(\kms)} } 
\startdata
1&                   RBS 38 &   M2.5V &   4.349247 & -66.753500 &  27.16 &  0.02 &  102.89 &  0.02 &  -16.83 &  0.02 &  10.30 &  0.40 &  1.0 \\ 
2&               GJ 2006 A &  M3.5Ve &   6.959886 & -32.551962 &  28.55 &  0.04 &  110.20 &  0.04 &  -47.19 &  0.04 &   8.80 &  0.40 &  1.8 \\ 
3&               HD 14082 B &     G2V &  34.353506 &  28.741465 &  25.23 &  0.02 &   85.88 &  0.02 &  -71.14 &  0.03 &   4.00 &  2.00 &  0.9 \\ 
4&                 AG Tri A &     K8V &  36.872305 &  30.973178 &  24.42 &  0.02 &   79.54 &  0.02 &  -72.13 &  0.01 &   6.00 &  0.30 &  0.9 \\ 
5&               CD-57 1054 &    K8Ve &  75.196664 & -57.256741 &  37.21 &  0.01 &   35.39 &  0.01 &   74.11 &  0.02 &  19.30 &  0.80 &  1.0 \\ 
6&                V1841 Ori &    K2IV &  75.205434 &  15.449933 &  18.83 &  0.02 &   18.31 &  0.02 &  -58.72 &  0.01 &  18.00 &  1.00 &  1.0 \\ 
7&             BD-21 1074 A &     M1V &  76.708210 & -21.585937 &  50.43 &  0.02 &   47.19 &  0.02 &  -15.38 &  0.02 &  21.00 &  2.00 &  1.3 \\ 
8&                   AO Men &   K3.5V &  94.617421 & -72.044516 &  25.57 &  0.01 &   -7.71 &  0.02 &   74.41 &  0.01 &  16.00 &  2.00 &  0.8 \\ 
9&           UCAC2 12510535 &    (M2) & 255.667191 & -45.366924 &  31.30 &  0.02 &  -20.10 &  0.02 & -137.85 &  0.02 &  -3.00 &  0.50 &  1.0 \\ 
10&              HD 164249 A &   F4.5V & 270.764224 & -51.649392 &  20.29 &  0.02 &    2.33 &  0.02 &  -86.23 &  0.02 &   0.20 &  0.40 &  1.1 \\ 
11&         UCAC4 299-160704 &    (M4) & 271.067437 & -30.308057 &  18.15 &  0.02 &    3.42 &  0.02 &  -65.22 &  0.02 &  -7.40 &  0.10 &  0.7 \\ 
12&  2MASS J18092970-5430532 &    (M4) & 272.373822 & -54.515278 &  25.67 &  0.03 &    4.71 &  0.03 & -108.17 &  0.02 &  -1.90 &  0.30 &  1.4 \\ 
13&    1RXS J184206.5-555426 &   M3.5V & 280.529082 & -55.907457 &  19.44 &  0.02 &   12.01 &  0.02 &  -79.07 &  0.01 &   0.60 &  0.20 &  1.3 \\ 
14&             Smethells 20 &    M1Ve & 281.719094 & -62.177194 &  19.72 &  0.02 &   13.24 &  0.02 &  -80.28 &  0.02 &   1.80 &  0.50 &  1.0 \\ 
15&              CD-31 16041 &   K8IVe & 282.685436 & -31.796817 &  20.22 &  0.01 &   17.27 &  0.02 &  -72.34 &  0.01 &  -8.00 &  3.00 &  1.0 \\ 
16&                HD 181327 &     F6V & 290.745786 & -54.538414 &  20.93 &  0.03 &   24.40 &  0.02 &  -82.19 &  0.02 &  -0.10 &  0.20 &  1.0 \\ 
17&    1RXS J192338.2-460631 &     M0V & 290.909327 & -46.109045 &  14.03 &  0.02 &   18.07 &  0.02 &  -57.25 &  0.01 &  -0.20 &  0.20 &  1.3 \\ 
18&         UCAC4 314-239934 &    (M2) & 297.068959 & -27.342442 &  15.47 &  0.02 &   25.15 &  0.02 &  -53.38 &  0.01 &  -6.26 &  0.05 &  1.3 \\ 
19&          TYC 7443-1102-1 &   K9IVe & 299.018392 & -32.127436 &  19.49 &  0.02 &   33.60 &  0.02 &  -68.53 &  0.01 &  -6.10 &  0.40 &  1.1 \\ 
20&         UCAC4 284-205440 &     M1V & 300.405042 & -33.220831 &  16.68 &  0.02 &   29.23 &  0.02 &  -61.39 &  0.01 &  -4.00 &  0.60 &  1.1 \\ 
21&                HD 191089 &     F5V & 302.271932 & -26.224333 &  19.96 &  0.02 &   40.34 &  0.02 &  -67.42 &  0.01 &  -7.00 &  1.00 &  0.9 \\ 
22&                   AU Mic &    M0Ve & 311.291183 & -31.342500 & 102.94 &  0.02 &  281.32 &  0.02 & -360.15 &  0.02 &  -4.50 &  0.60 &  0.9 \\ 
23&          GSC 06354-00357 &     M2V & 317.522749 & -19.333074 &  30.90 &  0.03 &   90.61 &  0.03 &  -90.99 &  0.02 &  -6.20 &  0.80 &  1.2 \\ 
24&              CPD-72 2713 &   K7IVe & 340.705235 & -71.706122 &  27.23 &  0.01 &   94.85 &  0.01 &  -52.38 &  0.01 &   7.80 &  0.60 &  0.9 \\ 
25&                   WW PsA &   M4IVe & 341.242468 & -33.251032 &  47.92 &  0.03 &  179.95 &  0.02 & -123.19 &  0.02 &   2.70 &  0.40 &  1.1 \\ 
26&                   HD 203 &     F3V &   1.709162 & -23.107750 &  25.16 &  0.03 &   96.97 &  0.04 &  -47.25 &  0.02 &   7.00 &  1.00 &  0.9 \\ 
27&                LP 525-39 &    M3Ve &   8.145477 &   7.490485 &  28.43 &  0.06 &  104.36 &  0.08 &  -62.91 &  0.06 &  -3.70 &  0.60 &  1.2 \\ 
28&                  GJ 3076 &      M5 &  17.856810 &  15.438867 &  58.00 &  7.30 &  192.00 &  8.00 & -130.00 &  8.00 &   2.50 &  0.30 & -- \\ 
29&             Barta 161 12 &    M4.3 &  23.808453 &  -7.214511 &  26.82 &  0.05 &   96.76 &  0.07 &  -48.94 &  0.03 &  10.00 &  4.00 &  1.5 \\ 
30&          PM J01538-1459A &     M3V &  28.462387 & -14.997647 &  29.64 &  0.04 &  106.72 &  0.04 &  -40.60 &  0.03 &   9.00 &  4.00 &  1.4 \\ 
\multicolumn{14}{c}{...}\\
\multicolumn{14}{c}{...}\\
\multicolumn{14}{c}{...}\\
   \enddata
   \tablecomments{The full machine-readable table is provided online.}       
\end{deluxetable*}

\section{Mean Group Sizes Using {\it inclusive members} \label{sec:appb}}

\begin{deluxetable}{lcc}
\tabletypesize{\small}
   \tablecaption{Mean Group Sizes Derived Through Three Distinct Methods \label{tab:size}} 
   \tablehead{ & \colhead{$-$Uncer} & \colhead{$+$Uncer} \\
       & \colhead{(pc)} & \colhead{(pc)}}
       \startdata
         \multicolumn{3}{c}{Method 1: Probabilistic Volume Calculation (PVC)} \\
       \hline
       \multicolumn{3}{c}{$bw$=7} \\
  %     \cmidrule(ll){3-3}
       $\rm RV_{obs}$ &  $-$ & 24.1$(^{+0.3}_{-0.3})$\\
       $\rm RV_{corr}$ &  $-$ & 25.7$(^{+0.3}_{-0.3})$ \\
       \multicolumn{3}{c}{$bw$=10} \\
 %      \cmidrule(ll){2-3}
       $\rm RV_{obs}$ & $-$ & 28.3$(^{+0.3}_{-0.3})$\\
       $\rm RV_{corr}$ & $-$ & 29.9$(^{+0.3}_{-0.3})$ \\
       \multicolumn{3}{c}{$bw$=15} \\
  %    \cline{3-3}
       $\rm RV_{obs}$  & $-$ & 35.9$(^{+0.2}_{-0.2})$\\
       $\rm RV_{corr}$ &  $-$ & 37.2$(^{+0.2}_{-0.2})$\\
       \multicolumn{3}{c}{$bw$=20} \\
  %     \cline{2-2}
       $\rm RV_{obs}$ & $-$ & 44.0$(^{+0.2}_{-0.2})$ \\
       $\rm RV_{corr}$ & $-$ & 45.3$(^{+0.2}_{-0.2})$ \\
        \hline       
        \multicolumn{3}{c}{Method 2: Mean Pairwise Distance Calculation} \\
        \hline
       $\rm RV_{obs}$ & 22.3$(^{+1.7}_{-1.9})$ & 24.8$(^{+2.0}_{-1.9})$ \\
       $\rm RV_{corr}$ & 24.9$(^{+1.8}_{-2.0})$ & 27.6$(^{+2.1}_{-2.1})$ \\
       \hline
       \multicolumn{3}{c}{Method 3: Covariance Matrix Calculation} \\
       \hline
       $\rm RV_{obs}$  & 8.4$(^{+0.6}_{-0.6})$ & 9.8$(^{+0.9}_{-0.8})$ \\
       $\rm RV_{corr}$ & 9.4$(^{+0.6}_{-0.7})$ & 10.9$(^{+1.0}_{-0.9})$ \\                       
\enddata
   \tablecomments{The results are obtained using the {\it inclusive members}. The reported uncertainties correspond to the 68\% confidence interval.}       
\end{deluxetable}

\section{Results Using {\it core members} \label{sec:appc}}

\begin{deluxetable}{lcc}
   \tablecaption{Ages for {\it core members} \label{tab:age_core}} 
   \tablehead{ & \colhead{$-$Uncer} & \colhead{$+$Uncer} \\
       & \colhead{(pc)} & \colhead{(pc)}}
       \startdata
        \multicolumn{3}{c}{Method 2: Mean Pairwise Distance Calculation} \\
        \hline
       $\rm RV_{obs}$ & 16.0$(^{+0.9}_{-1.3})$ & 14.8$(^{+1.6}_{-2.1})$ \\
       $\rm RV_{corr}$ & 18.9$(^{+1.6}_{-2.6})$ & 16.5$(^{+2.8}_{-4.2})$  \\
       \hline
       \multicolumn{3}{c}{Method 3: Covariance Matrix Calculation} \\
       \hline
       $\rm RV_{obs}$  & 16.4$(^{+1.1}_{-0.8})$ & 15.7$(^{+1.9}_{-1.7})$\\
       $\rm RV_{corr}$ &  20.6$(^{+4.8}_{-1.7})$ & 18.9$(^{+5.9}_{-2.9})$\\              
       \hline       
        \multicolumn{3}{c}{Summary\tablenotemark{a}} \\
       \hline
       $\rm RV_{obs}$  & 16.2$\pm$0.8 & 15.3$\pm$1.4 \\
       $\rm RV_{corr}$  & 19.3$\pm$2.3 & 17.3$\pm$3.4 \\       
\enddata
   \tablecomments{The reported uncertainties correspond to the 68\% confidence interval.}       
      \tablenotetext{a}{The summary ages were determined through a weighted mean of two methods. We utilized an age with $bw$=10 for the PVC, and we adopted larger uncertainties.} 
\end{deluxetable}

\begin{deluxetable}{lcc}
   \tablecaption{Mean Group Sizes for {\it core members} \label{tab:size_core}} 
   \tablehead{ & \colhead{$-$Uncer} & \colhead{$+$Uncer} \\
       & \colhead{(pc)} & \colhead{(pc)}}
       \startdata
        \multicolumn{3}{c}{Method 2: Mean Pairwise Distance Calculation} \\
        \hline
       $\rm RV_{obs}$ & 21.0$(^{+1.8}_{-2.0})$ & 23.1$(^{+2.2}_{-2.2})$ \\
       $\rm RV_{corr}$ & 23.3$(^{+1.9}_{-2.2})$ & 25.7$(^{+2.3}_{-2.4})$ \\
       \hline
       \multicolumn{3}{c}{Method 3: Covariance Matrix Calculation} \\
       \hline
       $\rm RV_{obs}$  & 7.4$(^{+0.6}_{-0.7})$ & 8.7$(^{+0.9}_{-0.9})$ \\
       $\rm RV_{corr}$ & 8.2$(^{+0.7}_{-0.8})$ & 9.7$(^{+1.1}_{-1.0})$ \\                   
\enddata
   \tablecomments{The reported uncertainties correspond to the 68\% confidence interval.}       
\end{deluxetable}

%% bibtext sample631.tex

\bibliography{kin_age_bpmg_rev2}{}
\bibliographystyle{aasjournal}



\end{document}